%% file: paper.tex
\documentclass{tlp}
\pdfoutput=1
\usepackage{amsmath,amssymb}
\usepackage{times,helvet,courier}
\usepackage{graphicx}
\usepackage{url}

\title{Multi-threaded ASP Solving with \textit{clasp}}
\author[Martin Gebser
  \and
  Benjamin Kaufmann
  \and
  Torsten Schaub]{%
  Martin Gebser
  and
  Benjamin Kaufmann
  and
  Torsten Schaub\thanks{Affiliated with Simon Fraser University, Canada, and Griffith University, Australia.}%
  \\
  Institut f\"ur Informatik, Universit\"at Potsdam}

\submitted{[n/a]}
\revised{[n/a]}
\accepted{[n/a]}

\begin{document}

\maketitle

\input{abstract}

\input{introduction}
\input{solving}

\input{architecture}

\input{communication}

\input{system}
\input{experiments}

\input{relwork}

\input{discussion}
%%% CAMERA
\input{acknowledgments}
\vspace{-0.35mm}
\input{bbl}
%\newpage
%\bibliographystyle{acmtrans}
%\bibliography{lit,akku,procs}
\end{document}

%% file: abstract.tex
\begin{abstract}
We present the new multi-threaded version of the 
state-of-the-art answer set solver \textit{clasp}.
We detail its component and communication architecture and illustrate how they
support the principal functionalities of \textit{clasp}.
Also, we provide some insights into the data representation used for
different constraint types handled by \textit{clasp}.
% \textit{clasp}'s several types of constraints.  
All this is accompanied by an extensive experimental analysis of the major
features related to multi-threading in \textit{clasp}.

\renewcommand\keywordsname{PUBLICATION NOTE}
{\keywords{ To appear in Theory and Practice of Logic Programming}}
\vspace{-1.5pt}
\end{abstract}

%%% Local Variables: 
%%% mode: latex
%%% TeX-master: "paper"
%%% End: 

%% file: introduction.tex
\section{Introduction}\label{sec:introduction}

The increasing availability of multi-core technology offers a great opportunity
for further improving the performance of solvers for Answer Set Programming
(ASP; \cite{baral02a}).
This paper describes how we % addressed this challenge in redesigning and reimplementing
redesigned and reimplemented
the award-winning%
\footnote{The multi-threaded variant of \textit{clasp~2} won 
  the first  place in the \emph{Crafted/UNSAT} and 
  the second place in the \emph{Crafted/SAT+UNSAT} category,
  respectively,
  at the \emph{2011 SAT competition} 
  in terms of number of solved instances and wall-clock time.
  In addition, 
        \textit{clasp~2} was among the three genuine parallel solvers participating in the
    \emph{32 cores track} (restricted to benchmarks from the \emph{Application} category;
 the fourth solver used a portfolio, including \textit{clasp~1.3}).
  Also, \textit{clasp~2} participated ``out of competition'' at the \emph{2011 ASP
 competition}, which was dominated by the single-threaded variant of \textit{clasp~2}.}
ASP solver \textit{clasp} \cite{gekanesc07a}
in order to leverage the power of today's multi-core shared memory % (CMP) 
machines
by supporting parallel search.
To this end,
we chose a coarse-grained, task-parallel approach via shared memory
multi-threading. 
This has led to the \textit{clasp~2} series supporting a single- and a
multi-threaded variant sharing a common code base.
\textit{clasp} allows for parallel solving 
by search space splitting and/or competing strategies.
While 
the former involves dynamic load balancing in view of highly irregular search
spaces,
both modes aim at running searches as independently as possible in order to take
advantage of enhanced sequential algorithms.
In fact, a portfolio of solver configurations cannot only be used for
competing   but also in splitting-based search.
The latter is optionally combined with global restarts to escape from uninformed
initial splits.

For promoting % supporting 
the scalability of parallel search, all major routines of \textit{clasp~2} are
lock-free.
Also, we enforced a clear distinction between read-only, shared, and
thread-local data and incorporated accordingly optimized representations.
This is implemented by means  of Intel's Threading Building Blocks (TBB) for
providing platform-independent threads, atomics, and concurrent containers.
Currently,
\textit{clasp} supports up to 64 configurable (non-hierarchic) threads.
Apart from parallel search,
another major extension of previous versions of \textit{clasp} regards the exchange of recorded % learned
nogoods.
While unary, binary, and ternary nogoods are always shared among all threads,
longer ones % and complex constraints 
can optionally be exchanged,
% The latter is 
configurable at the sender as well as at the receiver side.
% through options controlling the distribution and integration of nogoods.
In fact,
\textit{clasp} provides different measures estimating the quality of shared
nogoods as well as various heuristics and filters for controlling their
integration.
For instance, the sharing of a nogood can be subject to 
the number of distinct decision levels associated with its literals.
Conversely, 
the   integration of a nogood may depend on its satisfaction and/or scores in host heuristics.

In view of the wide distribution of \textit{clasp},
we put a lot of effort into transferring the entire functionality from the sequential,
viz.~\textit{clasp} series 1.3, to the parallel setting.
% To begin with, 
For one,       this concerned \textit{clasp}'s reasoning modes
(cf.\ \cite{gekakaosscsc11a}),
including
enumeration,
projected enumeration,
intersection and union 
of models,
and
optimization.
Moreover,
we extended \textit{clasp}'s language capacities by allowing for solving
weighted and/or partial MaxSAT \cite{liman09a}
as well as
% (weighted and/or partial) 
Boolean optimization \cite{maargrly11a} problems.
Finally, it goes without saying that \textit{clasp}'s basic infrastructure has also 
significantly evolved with the new design;
%Among others, 
e.g.\
the preprocessing capacities of \textit{clasp} were extended with blocked
clause elimination \cite{jabihe10a}, and
its conflict analysis has been significantly improved by on-the-fly
subsumption \cite{hansom09a}.

In what follows,
we focus on describing the multi-threaded variant of \textit{clasp~2}.
To this end, the next section provides a high-level view on modern parallel
ASP solving.
The general component and communication architecture of the new version
of \textit{clasp} are presented in Section~\ref{sec:architecture} and~\ref{sec:communication}.
Section~\ref{sec:system} details the design of data structures underlying the implementation of \textit{clasp~2}.
Parallel search features of \textit{clasp~2} are empirically assessed in Section~\ref{sec:experiments}.
Finally, Section~\ref{sec:relwork} and~\ref{sec:discussion} discuss
related work and the achieved results, respectively.

% \comment{To be rewritten after a full pass through paper}
% All in all, 
% the new design of \textit{clasp} has not only led to the addition of manifold
% new features,
% but moreover their interplay in the multi-threaded setting significantly
% increased the solving capacities of \textit{clasp}.
% In what follows,
% we focus on the multi-threaded architecture of \textit{clasp} and how it
% supports \textit{clasp}'s different reasoning capacities.
% Among them, new emerging feature include the exchange of learned constraints
% among solver threads as well as the parallelization of complex reasoning modes,
% like enumeration and optimization.
% Finally,
% we provide a first empirical evaluation of some selected features of
% \textit{clasp~2}.
% \com{T: Populate paper with references!}
%%% Local Variables: 
%%% mode: latex
%%% TeX-master: "paper"
%%% End: 

%% file: solving.tex
\newcommand{\skipline}{\vspace{-10pt}}
\newcommand{\skipsemi}{\vspace{-4pt}}
\newcommand{\skipsemii}{\vspace{-2pt}}
\newcommand{\skipsemiii}{\vspace{-3pt}}
\newcommand{\skipsemiiii}{\vspace{2pt}}

\section{Parallel ASP Solving}\label{sec:solving}

We presuppose some familiarity with search procedures for
(Boolean) constraint solving, that is,
Davis-Putnam-Logemann-Loveland (DPLL; \cite{davput60,dalolo62a})
and
Conflict-Driven Constraint Learning (CDCL; \cite{marsak99a,zamamoma01a}).
In fact, (sequential) ASP solvers like \textit{smodels} \cite{siniso02a}
adopt the search pattern of DPLL based on systematic chronological backtracking,
or like \textit{clasp} (series 1.3) apply lookback techniques from CDCL, which include
conflict-driven learning and non-chronological backjumping.
In what follows, we primarily concentrate on CDCL and % describe the 
principal points for its parallelization in the \textit{clasp~2} series.

In order to solve the basic decision problem of % whether there is a
solution existence,
CDCL first extends a given (partial) \emph{assignment} via deterministic (unit) propagation.
Importantly, every derived literal is ``forced'' by some \emph{nogood}
(set of literals that must not jointly be assigned), which would be violated
if the literal's complement were assigned.
Although propagation aims at forgoing nogood violations,
assigning a literal forced by one nogood may lead to the violation of another nogood;
this situation is called \emph{conflict}.
If the conflict can be resolved (the violated nogood contains backtrackable literals),
it is analyzed to identify a conflict constraint.
The latter represents a ``hidden'' conflict reason that is recorded and
guides backjumping to an earlier stage such that
the complement of some formerly assigned literal is forced by the conflict constraint,
thus triggering propagation.
Only when propagation finishes without conflict,
a (heuristically chosen) literal can be assigned at a new \emph{decision level},
provided that the assignment at hand is partial,
while a \emph{solution} (total assignment not violating any nogood)
has been found otherwise.
The eventual termination of CDCL is guaranteed (cf.\ \cite{zhamal03a,ryan04a}),
by either returning a solution or encountering an unresolvable conflict
(independent of unforced decision literals).

% ------------------------------------------------------------
\begin{figure}[t]
\makebox[\textwidth]{\hrulefill}
\raggedright\hspace*{1em}
\textbf{while} work available
\skipsemi%
\begin{itemize}
\item [] \textbf{while} no (result) message to send % no messages
\skipsemii%
  \begin{itemize}
  \item [] \textit{communicate}
    \hfill// exchange information with other solver instances%
  \item [] \textit{propagate}  
    \hfill// deterministically assign literals% compute deterministic consequences
\skipsemiiii%
  \item [] \textbf{if} no conflict \textbf{then}
\skipsemii%
    \begin{itemize}
    \item [] \textbf{if} all variables assigned 
      \textbf{then} 
      \textbf{send} solution % variable assignment
    \item [] \textbf{else}
      \textit{decide} 
      \hfill// non-deterministically assign some literal%
\skipsemii%
    \end{itemize}
  \item [] \textbf{else} 
\skipsemii%
    \begin{itemize}
    \item [] \textbf{if} root-level conflict
      \textbf{then} 
      \textbf{send} unsatisfiable
    \item [] \textbf{else if} external conflict
      \textbf{then} 
      \textbf{send} unsatisfiable
    \item [] \textbf{else}
\skipsemi%
      \begin{itemize}
      \item [] \textit{analyze} \hfill// analyze conflict and add conflict constraint%
\skipsemiii%
      \item [] \textit{backjump} \hfill// % undo assignments 
                                          unassign literals until conflict constraint is unit%
%\skipsemii%
      \end{itemize}
    \end{itemize}
  \end{itemize}
% \item [] 
  \textit{communicate}
  \hfill// exchange results with (and receive work from) other solver instances%
\end{itemize}%
\skipline%
\makebox[\textwidth]{\hrulefill}%
\skipsemi\skipsemii%
\caption{% Decision 
         High-level algorithm for multi-threaded Conflict-Driven (Boolean) 
         Constraint Learning.} % (CDCL)}
\label{fig:cdcl}
\end{figure}
% ------------------------------------------------------------
%
% The procedure in 
Figure~\ref{fig:cdcl} provides a high-level view on the
parallelization of CDCL-style search in \textit{clasp}.
% To begin with, 
We first
note that entering the inner search loop relies on the
availability of work.
In fact, when search spaces to investigate in parallel are split up by means of
\emph{guiding paths} \cite{zhbohs96a},
a solver instance must        acquire some spare guiding path before it can start to search.
In this case, all (decision) literals of the guiding path are assigned up to the solver's
\emph{root level}, precluding them from becoming unassigned upon backtracking/backjumping.
Apart from search space splitting,
parallelization of \textit{clasp} can be based on \emph{algorithm portfolios} \cite{gomsel01a},
running different solving strategies competitively on the same search space.
Once a solver instance is working on some search task,
it combines deterministic propagation with communication.
The latter includes nogood exchange with other solver instances,
work requests from idle solvers (asking for a guiding path),
and external conflicts raised to abort the current search.\footnote{%
For instance, a solver instance may discover unconditional unsatisfiability
(even when using guiding paths; cf.\ \cite{elgegukakaliscscsc09a})
and then inform others about the needlessness of performing further work.}
An external conflict or an (unresolvable) root-level conflict
likewise make a solver instance stop its current search, and the same
applies when a solution is found.
In such a case, the respective result is communicated
(in the last line of Figure~\ref{fig:cdcl}),
and a new search task may be received in turn.

As mentioned in the introductory section,
the infrastructure of \textit{clasp} also allows for
conducting sophisticated reasoning modes like
enumeration and optimization in parallel.
This is accomplished via enriched message protocols,
e.g.\ (upper) bounds are exchanged in addition to nogoods 
when performing parallel optimization,
while an external conflict (raised upon finding the first solution)
switches competing solvers of an algorithm portfolio into
enumeration mode based on guiding paths.
In fact, search space splitting and algorithm portfolios can be
applied exclusively or be combined to flexibly orchestrate parallel solvers.

In the following sections,
we detail the parallel architecture and underlying
implementation techniques of \textit{clasp~2}.
Regarding data structures,
it is worthwhile to note that unit propagation over ``long'' nogoods
(involving more than three literals)
relies on a \emph{two-watched-literals} approach \cite{momazhzhma01a},
monitoring two references to unassigned literals for triggering propagation
once the second last literal becomes assigned.
We also presuppose basic familiarity with parallel computing concepts,
such as race conditions, atomic operations, (dead- and spin-) locks,
semaphores, etc.\ (cf.\ \cite{hersha08a}).

% \com{Background}
% In what follows,
% we presuppose some familarity with the theory and practise of ASP.
% \begin{itemize}
% \item nogoods
% \item assignment
% \item decicion level
% \item guiding path

% use the well-known \emph{guiding path}
% technique~\cite{zhbohs96a} for splitting the search space into disjoint parts.
% \item parallel algorithm portfolios~\cite{gomsel01a}.
% \item watched literals
% \end{itemize}

% \begin{itemize}
% \item atomic datastructure
% \item lock, deadlocks, semaphores, races, spinlocks
% \end{itemize}

% % \tbrw

% \begin{itemize}
% \item decision problem
%   \begin{itemize}
%   \item guiding path/search space partition
%   \item portfolio/competition
%   \item guiding path/search space partition/portfolio
%   \end{itemize}
% \item enumeration
%   \begin{itemize}
%   \item interplay of competition and partition
%   \end{itemize}
% \item optimization
%   \begin{itemize}
%   \item interplay of competition and partition
%   \end{itemize}
% \end{itemize}

%%% Local Variables: 
%%% mode: latex
%%% TeX-master: "paper"
%%% End: 

%% file: architecture.tex
\section{Component Architecture}\label{sec:architecture}

\begin{figure}[t]
  \centering
  \includegraphics[height=.35\textheight]{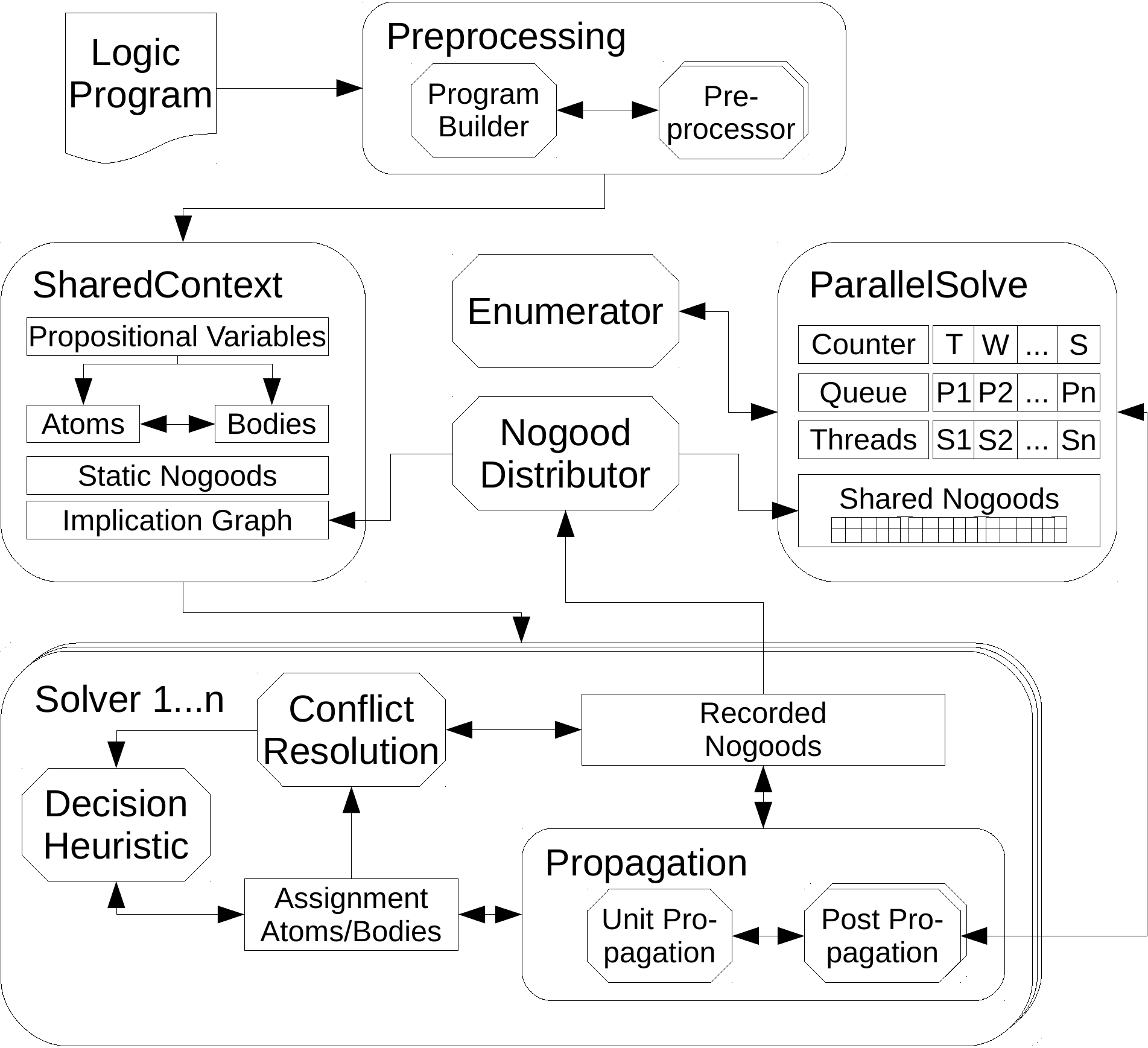}%
\skipsemi\skipsemii%
  \caption{Multi-threading architecture of \textit{clasp~2}.}
  \label{fig:arch}
\end{figure}

To explain the architecture and functioning of the new version of \textit{clasp},
let us follow the workflow underlying its design.
To this end,
consider \textit{clasp}'s architectural diagram given in Figure~\ref{fig:arch}.
% ------------------------------------------------------------
% ------------------------------------------------------------
Although \textit{clasp} also accepts other input formats, 
like (extended) \textit{dimacs}, \textit{opb}, and \textit{wbo} for
describing % (Maximum) 
                     Boolean satisfiability (SAT; \cite{SATHandbook})
%                        and (Pseudo- or Weighted) Boolean 
                         and                               optimization problems,
% \comment{T: mention new formats}
we detail its functioning for computing answer sets of (propositional) logic programs,
as output by grounders like \textit{gringo} \cite{gekakaosscsc11a} or \textit{lparse} \cite{lparseManual}.
Similarly, we concentrate on the multi-threaded setting, neglecting the
single-threaded one.

At the start,
only the main thread is active.
Once the logic program is read in,
it is subject to several preprocessing stages,
all conducted by the main thread.
At first, the program is (by default) simplified while identifying equivalences
among its constituents \cite{gekanesc08a}.
The simplified program is then transformed into a compact representation in
terms of Boolean constraints 
(whose core is generated from the completion \cite{clark78} of the simplified program).
After that,
the constraints are (optionally) subject to further, mostly SAT-based
preprocessing \cite{eenbie05a,jabihe10a}.
Such techniques are more involved in our ASP setting  because variables
relevant to unfounded-set checking,  optimization, or part of complex (i.e.\ cardinality and weight)
constraints cannot be simply eliminated.
Note that both preprocessing steps identify  redundant variables that can be
expressed in terms of the relevant ones  included in the resulting set of
constraints.

The outcomes of the preprocessing phase are stored in a
{\small\textsf{SharedContext}} object
that is initialized by the main thread and shared among all
participating threads.
% The contents of this object is shared among all
% participating threads.
% Among others, it contains
Among others, this object contains
\begin{itemize}
\item the set of relevant Boolean variables together with type information\\
  (e.g.\ atom, body, aggregate, etc.),
\item a symbol table, mapping (named) atoms from the program to internal variables,
\item the positive atom-body dependency graph,
  restricted to its strongly connected components,
\item the set of Boolean constraints, among them 
  nogoods, cardinality and weight constraints, minimize constraints,
  and
\item an implication graph capturing inferences from binary and ternary nogoods.%
  \footnote{ASP problems usually yield a large majority of binary nogoods due to
  program completion \cite{clark78}.
  Also note that unary nogoods capture initial problem simplifications
  that need not be rechecked during search.\label{foo:binary}}
% \comment{unary?}
\end{itemize}
The richness of this information is typical for ASP, and it
is much sparser in
a SAT setting.

% The {\small\textsf{SharedContext}} object also hosts the master solver.
% Upon initialization,
% this solver takes the initial set of constraints and performs
% solver-specific simplifications.
% Once this is done,
After its initialization in association with a ``master solver,''
further (solver) threads are (concurrently) attached to the
{\small\textsf{SharedContext}}, where its constraints are ``cloned.''
Notably, each constraint % knows 
is aware of how to clone itself efficiently
(cf.\ Section~\ref{sec:system} on implementation details).
Moreover, the % {\small\textsf{SharedContext}} hosts the global
{\small\textsf{Enumerator}} and {\small\textsf{NogoodDistributor}} objects are
used globally in order to coordinate % for 
various model enumeration modes and nogood exchange among % the solvers.
solver instances.
We detail their functioning in Section~\ref{sec:communication}.

Each thread contains one {\small\textsf{Solver}} object, implementing the
algorithm in Figure~\ref{fig:cdcl}.
Each {\small\textsf{Solver}} stores
\begin{itemize}
\item local data,       \ including assignment, watch lists, constraint database, etc.,
\item local strategies, \ regarding heuristics, restarts, constraint deletion, etc.,
\end{itemize}
and it uses the {\small\textsf{NogoodDistributor}} to share recorded nogoods.
A solver assigns variables either by (deterministic) propagation or
(non-deterministic) decisions.
Motivated by the nature of ASP problems,\footnotemark[3] %[\ref{foo:binary}]
each solver propagates
first binary and ternary nogoods (shared through the aforementioned implication graph),
then longer nogoods
and
other constraints,
before it finally applies any available post propagators.

Post propagators constitute another important new feature of \textit{clasp~2},
providing an abstraction easing \textit{clasp}'s extensibility with more
elaborate propagation mechanisms.
For this,
each solver maintains a list of post propagators that are consecutively
processed after unit propagation.
For instance,
failed-literal detection and unfounded-set checking are implemented in \textit{clasp~2} as post
propagators.
Similarly, 
they are used in the new version of \textit{clasp}'s extension with constraint processing, 
\textit{clingcon} \cite{geossc09a}, to realize theory propagation.
Post propagators are assigned different priorities and are called in priority
order. 
Typically, we distinguish three priority classes:
\begin{itemize}
\item \textit{single} post propagators are deterministic and only extend the current decision level.\\
  Unfounded-set checking is a typical example.
\item \textit{multi} post propagators are deterministic and may add or remove decision levels.\\
  Failed-literal detection is a typical example.
\item \textit{complex} post propagators may or may not be deterministic.\\
  Nogood exchange is an example for this (see below).
\end{itemize}
Moreover, parallelism is also handled by means  of post propagators,
as described next.

{\small\textsf{ParallelSolve}} controls % parallel 
concurrent solving 
with up to 64 individually configurable threads.
When attaching a solver to the {\small\textsf{SharedContext}},
{\small\textsf{ParallelSolve}} associates a thread with the solver and adds
dedicated post propagators to it. % each solver.
One high-priority post propagator is added for message handling
and another, very low-priority post propagator is supplied
for integrating information stemming from models%
\footnote{This can regard an enumerated model to exclude, intersect, or union, 
  as well as objective function values.} 
and/or shared nogoods.

For controlling parallel search, {\small\textsf{ParallelSolve}} maintains a set
of atomic message flags:
\begin{itemize}
\item \textit{terminate} signals the end of a computation,
\item \textit{interrupt} forces outside termination (e.g.\ when the user hits Ctrl+C),
\item \textit{sync}      indicates that all threads shall synchronize, and
\item \textit{split}     is set during splitting-based search whenever at least one thread needs work.
\end{itemize}
These flags are used to implement \textit{clasp}'s two major search strategies:
\begin{itemize}
\item \emph{splitting-based search} via distribution of guiding paths and
  dynamic load balancing via a split-request and -response protocol,
  and 
\item \emph{competition-based search} via freely configurable solver portfolios.
\end{itemize}
Notably, solver portfolios can also be used in splitting-based search, that is,
different guiding paths may be solved with different configurations.

% Finally,
% we just mention the components {\small\textsf{Enumerator}} and
% {\small\textsf{NogoodDistributor}}.
% While the former is in charge of implementing all enumeration-based reasoning
% modes of \textit{clasp~2},
% the latter handles the exchange of recorded nogoods among solver threads.
% Both components deal with communication-intense processes and are thus
% detailed in the next section.

%%% Local Variables: 
%%% mode: latex
%%% TeX-master: "paper"
%%% End: 

%% file: communication.tex
\section{Communication Architecture}
\label{sec:communication}

A salient transverse aspect of the architecture of \textit{clasp~2} is its
communication infrastructure, used for implementing advanced reasoning
procedures.
To begin with, the {\small\textsf{ParallelSolve}} object keeps track of
threads' load, particularly in splitting-based search.
Moreover,
the {\small\textsf{Enumerator}} controls enumeration-based reasoning modes,
while
the {\small\textsf{NogoodDistributor}} handles the exchange of recorded nogoods among solver threads.
These communication-intense components along with fundamental implementation techniques
are detailed below in increasing order of complexness.

\subsection{Thread Coordination}

The basic communication architecture of \textit{clasp} relies on message passing,
efficiently implemented by lock-free atomic integers.
On the one hand, globally shared atomic counters are stored in {\small\textsf{ParallelSolve}}.
For instance, all aforementioned control flags are stored in a single shared atomic integer.
On the other hand, each thread has a local message counter hosted by the message handling post
propagator (see above).
Message passing builds upon two basic methods:
\texttt{postMessage\!()} and \texttt{hasMessage\!()}.
Posting a message amounts to a \emph{Compare-And-Swap}%
\footnote{Conditional writing is performed as atomic CPU instruction 
          to achieve synchronization in multi-threading.}
(CAS)
on an atomic integer,
and checking for messages (via specialized post propagators) is equivalent to
an  atomic read.
Of particular interest is communication during splitting-based search.
This is accomplished via
a lock-free work queue,
an atomic work request counter,
and
a work semaphore
in {\small\textsf{ParallelSolve}}.
Initially, the work queue only contains the empty guiding path, and all
threads ``race'' for this work package by issuing a work request.
A work request first tries to pop a guiding path from the work queue and returns upon success.
Otherwise, 
the work request counter is incremented and a split request is posted, 
which results in raising the \textit{split} flag.
Afterwards, a {\texttt{wait\!()}}
is tried on the work semaphore.%
\footnote{See \url{http://en.wikipedia.org/wiki/Semaphore_(programming)} in case
  of unfamiliarity with the working of semaphores.}
If {\texttt{wait\!()}} fails because the number of idle threads now equals
the total number of threads, the requesting thread posts a \emph{terminate}
message and wakes up all waiting threads.  
Otherwise, the thread is blocked until new work arrives.
On the receiver side,
the message handling post propagator of each thread checks whether the
\textit{split} flag has been set.
If so, and provided that the thread at hand has work to split,
its message handler proceeds as follows.
At first, it decrements the work request counter.
(Note that the message handler thus declares the request as handled before
actually serving it in order to minimize over-splitting.)
If the work request counter reached 0, the message handler also resets the \textit{split} flag.
Afterwards, the search space is split and a (short) guiding path is pushed to
the work queue in {\small\textsf{ParallelSolve}}.
At last, the message handler signals the work semaphore and hence eventually wakes up a waiting thread.

Splitting-based search usually suffers from uninformed early splits of the
search space.
To counterbalance this, {\small\textsf{ParallelSolve}} supports an advanced
global restart scheme based on a two-phase strategy.
In the first phase, threads vote upon effectuating a global restart based on
some given criterion (currently, number of conflicts);
however, individual threads may veto a global restart.
For instance, this may happen in enumeration when a first model is found during
this first restarting phase.
Once there are enough votes,
a global restart is initiated in the second phase.
For this, a \textit{sync} message is posted and 
threads wait until all solvers have reacted to this message.
The last reacting thread decides on how to continue.
If no veto was issued, the global restart is executed. That is, threads
give up their guiding paths, the work queue is cleared, 
and the initial (empty) guiding path is again added to the work queue.
Otherwise, 
the restart is abandoned, and the threads simply continue with their current guiding paths.

If splitting-based search is not active (i.e.\ during competition-based search),
the work queue initially contains one (empty) guiding path for each thread, and additional
work requests simply result in the posting of a \emph{terminate} message.

\subsection{Nogood Exchange}\label{subsec:nogood}

Given that each thread implements conflict-driven search involving nogood learning,
the corresponding solvers may benefit from a controlled exchange of their
recorded information.
However, such an interchange must be handled with great care because each
individual solver may already learn exponentially many nogoods,
so that their additional sharing % distribution 
may significantly hamper the overall
performance.

% To this end,
To differentiate which nogoods to share,
% we pursue in 
\textit{clasp~2} pursues a hybrid approach regarding both nogood exchange and storage.
% We have already seen in Section~\ref{sec:architecture} that 
As described
in Section~\ref{sec:architecture}, the binary and
ternary implication graph (as well as the positive atom-body dependency graph) are shared
among all solver threads.
% In this way,
% all threads share all binary and ternary nogoods as well as implicitly all loop nogoods.
Otherwise, each solver maintains its own local nogood database.
The sharing of these nogoods is optional, as we detail next.

The actual exchange of nogoods is controlled in \textit{clasp} by separate 
distribution and integration components 
for carefully selecting the spread constraints.
This is supported by thread-local interfaces along with the global
{\small\textsf{NogoodDistributor}} (see Figure~\ref{fig:arch}).
All components rely on interfaces abstracting from the specific sharing % distribution
mechanism used underneath. % for the different types of constraints. % (see below).

The distribution of nogoods is configurable in two ways.
First, the exported nogoods can be filtered by their \emph{type},
viz.\ {conflict}, {loop}, or {short} (i.e.\ binary and ternary),
or be exhaustive or inhibited.
The difference between globally sharing short nogoods (via their implication graph) 
and additionally ``distributing'' them lies in the proactiveness of the process.
While the mere sharing leaves it to each solver to discover nogoods added by
others,
their explicit distribution furthermore % actively 
communicates this information
through the standard distribution process.
Second, the export of nogoods is subject to their respective number of distinct
decision levels associated with the   contained literals,
called the \emph{Literal Block Distance} (LBD; \cite{audsim09a}).
Fewer distinct decision levels are regarded as advantageous since they
are prone to prune larger parts of the search space.
This criterion has empirically shown to be rather effective and largely superior
to a selection by length.

% Similarly t
The integration of nogoods is likewise configurable in two ways.
The first criterion captures the \emph{relevance} of a nogood to the local search
process.
First, the state of a nogood is assessed by checking whether it is
satisfied,
violated,
open (i.e.\ neither satisfied nor violated),
or
unit w.r.t.\ the current (partial) assignment.
While violated and unit nogoods are always considered relevant, 
open nogoods are optionally passed through a filter using the solver's current
heuristic values to discriminate the relevance of the candidate nogood to the
current solving process.
Finally, satisfied nogoods are either ignored or considered open
depending on the configuration of the corresponding filter and
their state relative to the original guiding path.
The second integration criterion is expressed by a \emph{grace period}
influencing the size of the local import queue and thereby the minimum time a
nogood is stored. % retained.
Once the local import queue is full,
the least recently added nogood is evicted and either 
transferred to the thread's nogood database (where it becomes subject to the thread's nogood
deletion policy) or immediately discarded.
Currently, two modes are distinguished.
The thread transfers either all or only ``heuristically active'' nogoods from
its import queue while discarding all others. 

Both distribution and integration are implemented as dedicated (complex) post propagators,
% Their current implementation is 
% These post propagators are 
based upon a global distribution scheme
implemented via
an efficient lock-free \emph{Multi-Read-Multi-Write}
(MRMW) % \footnote{MRMW stands for \emph{Multi Read Multi Write}.}
list situated in {\small\textsf{ParallelSolve}}.%
\footnote{This choice is motivated by the fact that we aim at optimizing
  \textit{clasp} for desktop computers, still mostly possessing few genuine processing units.
  Other strategies are possible and an active subject of current research.}
Distribution roughly works as follows.
When the solver of Thread~$i$ records a nogood that is a candidate for sharing,
it is first % of all 
integrated into % its 
the thread-local nogood database.
In addition, the nogood's reference counter is set to the total number of threads plus one,
and its target mask to all threads except~$i$.
At last,
Thread~$i$ appends the shared nogood to the aforementioned MRMW list.

Conversely upon integration,
Thread~$j$ traverses the MRMW list, thereby ignoring all nogoods 
whose target mask excludes~$j$.
Depending on the state of a nogood, 
the aforementioned filters decide whether a nogood is relevant or not.
All relevant nogoods are integrated into the search process of Thread~$j$
% \comment{How to integrate nogoods into the search? Reference to implementation?}
and added to its local import queue. 
The reference counter of each nogood is decremented by each thread moving its
read pointer beyond it.
In addition, the sharing thread~$i$ decrements a nogood's reference counter
whenever it no longer uses it.
Hence, the reference counter of a shared nogood can only drop to zero
once it is no longer addressed by any read pointer.
This makes it subject to deletion.

Notably,
the shared representation of a nogood is only created when the nogood is 
actually distributed. 
Otherwise, its optimized (single-threaded) representation is used.
Upon integration, the ``best'' representation is selected, for instance,
short nogoods are copied while longer ones are physically shared
(see Section~\ref{sec:system} for implementation details).

\subsection{Complex Reasoning Modes}

In addition to model printing,
all enumeration-based reasoning modes of \textit{clasp~2} are controlled by the
global {\small\textsf{Enumerator}} (see Figure~\ref{fig:arch}).
These reasoning modes include
regular and projected model enumeration,
intersection and union of models,
uniform and hierarchical (multi-criteria) optimization
as well as combinations thereof, 
like computing the intersection of all optimal models.

As already mentioned,
one global {\small\textsf{Enumerator}} is shared among all threads and is
protected by a lock.
Whenever applicable, it hosts global constraints, like minimize constraints,
that are updated whenever a model is found.
Additionally, the {\small\textsf{Enumerator}} adds a local enumeration-specific
constraint to each solver for storing thread-local data, 
e.g.\ current optima (see below).
Once a model is found,
a dedicated message \textit{update-model} is send to all threads,
but threads only react to the most recent one.

In fact, enumeration is combinable with both search strategies described in
Section~\ref{sec:architecture},
either by applying  dedicated enumeration algorithms taking advantage of guiding
paths or by using solution recording in a competitive setting.
The latter setting exploits the infrastructure for nogood exchange in
order to distribute solutions among solver threads.
Once a solution is converted into a nogood, it can be treated as usual,
except that its integration is imperative and that it is exempt from deletion.
However, this approach suffers from exponential space complexity in the worst case.
Unlike this, splitting-based enumeration runs in polynomial space,
following a distributed version of the enumeration algorithm introduced
in \cite{gekanesc07c}.
In order to avoid uninformed splits at the beginning,
all solver threads may optionally start in a competitive setting.
Once the first model is found,
the {\small\textsf{Enumerator}} enforces splitting-based search among all solver
threads and disables global restarts.
In addition to the distribution of disjoint guiding paths,
backtrack levels (see \cite{gekanesc07c}) are dealt with locally in order to
guarantee an exhaustive and duplicate-free enumeration of all models.

In optimization, solver threads cooperate in enumerating one better model after
another until no better one is found,
so that the last model is optimal. % an optimum one.
Whenever a better model is found,
its objective value is stored in the {\small\textsf{Enumerator}}.
The threads react upon the following \textit{update-model} message by
integrating the new value into their local
% optimization-specific constraint, viz.\ a minimize constraint, and with it into the solver's search process.
minimize constraint representation%
\footnote{%
While the literals of a minimize constraint are stored globally,
% in the {\small\textsf{Enumerator}},
corresponding upper bounds are local to threads,
and changes are communicated through the {\scriptsize\textsf{Enumerator}}.}
and thus into the search processes of their solvers.
Minimize constraints provide methods for efficiently re-computing their state
after an update, so that restarting search is unnecessary in most cases.
An innovative feature of \textit{clasp~2} is hierarchical optimization \cite{gekakasc11c},
build on top of uniform optimization.
Hierarchical optimization allows for solving multi-criteria optimization
problems by considering criteria according to their respective priorities.
Such an approach is much more involved than standard branch-and-bound-based
optimization because it %s search process 
must recover from several unsatisfiable
subproblems, one for each criterion.
This is accomplished by dynamic minimize constraints that may be disabled and
reinitialized during search.
Accordingly, nogoods learned under minimize constraints must be retracted once
the constraint gets disabled.
Another benefit of such dynamic constraints is that we may decrease the (upper) bound
in a non-uniform way, and successively re-increase it upon unsatisfiability.
Hierarchical optimization allows for gaining an order of magnitude on
multi-criteria problems, as witnessed in Linux configuration \cite{gekakasc11d}.

Also, brave and cautious reasoning, computing the union and intersection of all
models, respectively, are implemented through a global constraint within the
{\small\textsf{Enumerator}}.
Whenever a new model is found,
the constraint is intersected
% or unioned,
% respectively, with the model's complement.
with the model (or its complement).

%%% Local Variables: 
%%% mode: latex
%%% TeX-master: "paper"
%%% End: 

%% file: system.tex
\section{Implementation}\label{sec:system}

A major design goal of \textit{clasp~2} was to leverage the power of today's
multi-core shared memory machines,
while keeping the resulting overhead low so that the single-threaded variant
does not suffer from a significant loss in performance.
In particular,
we aimed at empowering physical sharing of constraints and data 
while avoiding false sharing, locking, and communication overhead.
To this end,
our design foresees a clear distinction between three types of data representations,
viz.
\begin{itemize}
\item \textit{read-only} % shared 
  data providing lock- and wait-free sharing %,
  (without deadlocks and races),
\item \textit{shared} data being subject to concurrent updates via CAS or locks
  (admitting races), and
\item \textit{thread-local} data being private to each thread and thus not sharable
  (avoiding deadlocks and races).
\end{itemize}
Let us make this more precise by detailing the data representations of the
various types of constraints used in \textit{clasp}.
Constraints are typically separated into a thread-local and a (possibly shared)
read-only part.
While the former usually contains search-specific and thus dynamic data,
the latter typically comprises static data not being subject to change.

As mentioned above, the % (binary and ternary) 
\textbf{implication graph} is shared
among all threads and stores inferences from binary and ternary nogoods.
The corresponding data structure is separated into two parts. 
On the one hand,
a static read-only part is initialized during preprocessing;
it stores two vectors, \texttt{bin\!(l)} and \texttt{tern\!(l)},
for each literal \texttt{l}. 
The former contains literals being forced once \texttt{l} becomes true.  
Similarly, the latter stores binary clauses being % implied 
activated when \texttt{l} becomes true. 
For better data locality, \texttt{bin\!(l)} and \texttt{tern\!(l)} are actually
stored in one memory block.
On the other hand,
the dynamic part supports % efficient lock-free 
concurrent updates for 
storing and distributing short recorded nogoods. % learned clauses.
To this end,
it includes, for each literal~\texttt{l}, an atomic pointer, \texttt{learnt\!(l)}, 
to a linked list of \texttt{CACHE\_LINE\_SIZE}-sized memory blocks.  
Each such memory block contains   a fixed-size array of binary and ternary nogoods.
This setting guarantees that propagation over \texttt{learnt\!(l)} is efficient
and does not need any locks (given that short clauses are never removed).
Moreover, we rely on fine-grained spinlocks to enable efficient updates of
fixed-size arrays.
%
% (Recall from Section~\ref{sec:architecture} that the binary and ternary nogoods are always propagated first.)

In analogy, longer \textbf{nogoods} are separated into two parts, called head and tail.
The head part is always thread-local and is referenced in the owning thread's watch lists.
It stores two watched literals, one cache
literal, and some extra dynamic data, like nogood activity.
The cache literal provides a (potential) spare watched literal,
in case one of the two original ones is assigned.
That is, upon updating the watched literals,
the cache literal is inspected before a costly visit of the literals in the
(possibly shared) tail part is engaged.\footnote{%
The \emph{Watched Literal Reference Lists} of \textit{miraxt} \cite{sclebe09a} follow a similar approach.
} 
Further contents of the head part depend on whether a   nogood is shared.
If not,
the nogood stores its unshared tail part,
including the nogood's size and remaining literals,
together with the head in one continuous memory block. 
Otherwise, the head points to a read-only shared tail object containing
the nogood's literals, an (atomic) reference counter, and further static data, like the
size of the nogood.
The separation into a dynamic thread-local and a static read-only shared part is
motivated by the fact that sharing only needs to replicate the search-specific
state of a nogood, like its watched literals and activity.
Notably,
although a more local representation of shared nogoods would be possible,
it is important to avoid storing dynamic data of different threads 
in the same coherence block (e.g.\ a cache line);
otherwise, writes of one
thread lead to (logically) unnecessary coherence operations in other threads.
Our separation of data ensures that thread-local data of different threads 
is never stored together and thus avoids such ``false sharing.''
Regarding representation, \textit{clasp} employs the following policies.
Short nogoods of up to five literals are never physically shared,
but completely stored in thread-local head parts for improving access locality.
Original problem nogoods are physically shared in the presence of
multiple threads,
except if copying (instead of sharing) of problem nogoods is enforced.
Finally, recorded nogoods are only shared on demand, 
as described in Section~\ref{sec:communication}.

Analogously to nogoods,
\textbf{weight constraints} have a
thread-local part storing current assignments 
(to enclosed literals) and the corresponding sum of weights as well as 
a shared part storing size, literals, weights, and a reference counter.
The shared part of a \textbf{minimize constraint} (cf.\ Section~\ref{sec:communication})
in addition includes priority levels of literals,
and thread-local parts contain current (upper) bounds.

Finally,
\textbf{unfounded-set checking} also relies on a bipartite data representation.
As mentioned above, it is implemented as a dedicated post propagator
utilizing the (read-only) shared strongly connected components of a program's
positive atom-body dependency graph (cf.\ Section~\ref{sec:architecture}).
% As above, 
This is again counterbalanced by a thread-local part storing
assignment-specific data, like source pointers (cf.\ \cite{siniso02a}).

%%% Local Variables: 
%%% mode: latex
%%% TeX-master: "paper"
%%% End: 

%% file: experiments.tex
\newcommand{\itgap}{\hspace{0.7pt}}

\section{Experiments}\label{sec:experiments}

We conducted two series of experiments, the first comparing \textit{clasp~2}
to other multi-threaded CDCL-based (SAT) solvers and the second assessing the impact of
different parallel search features.
In fact, efforts to parallelize CDCL have so far concentrated on the area of SAT,
and thus we compare \textit{clasp} (version~2.0.5) to the following multi-threaded SAT solvers:
\textit{cryptominisat} (version~2.9.2; \cite{sonoca09a}),
\textit{manysat} (version~1.1; \cite{hajasa09c}),
\textit{miraxt} (version~2009; \cite{sclebe09a}), and
\textit{plingeling} (version~587f; \cite{biere11a}).
While \textit{miraxt} performs search space splitting via guiding paths,
the three other solvers let
different configurations of an underlying sequential SAT solver 
compete with one another.
Furthermore, nogood exchange among individual threads is either confined to short nogoods,
only unary (\itgap\textit{plingeling}) or binary ones as well (\textit{cryptominisat}),
performed adaptively (\textit{manysat}; cf.\ \cite{hajasa09b}), or
exhaustive in view of a shared nogood database (\textit{miraxt}).
% All solvers were run (exclusively) on a machine 
The solvers were run on a Linux machine 
% equipped 
with two Intel Quad-Core Xeon E5520 2.27GHz processors, % under Linux,
imposing a limit of 1000 (or 1200) seconds wall-clock time per solver and benchmark instance
in the first (or second) series of experiments.\footnote{%
 The benchmark suites are available at
 \url{http://www.cs.uni-potsdam.de/clasp}.}
% ------------------------------------------------------------
\begin{figure}[t]
  \centering%\hspace*{-8mm}
  \includegraphics[width=0.5\textwidth]{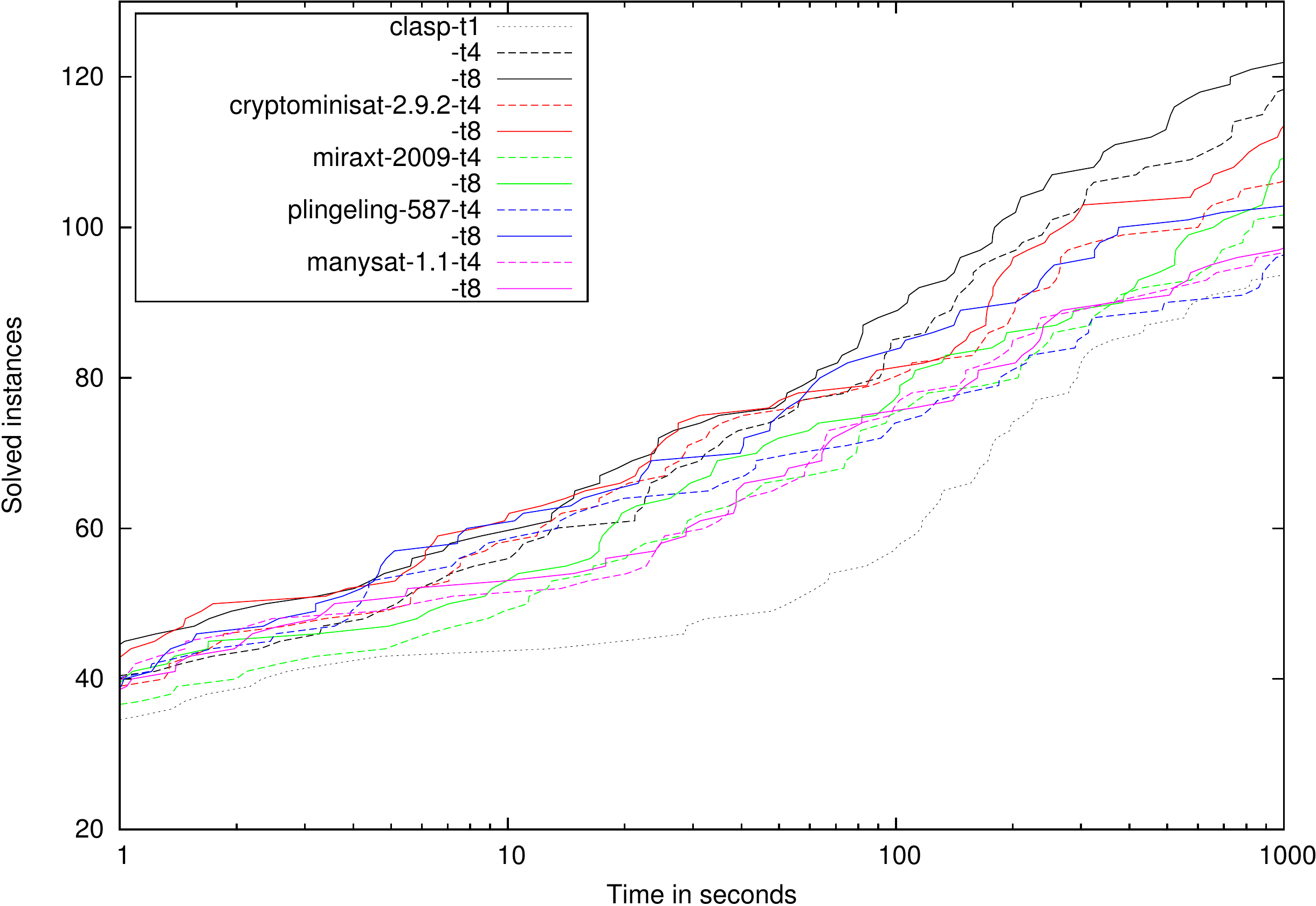}%
\skipsemi\skipsemii%
  \caption{Number of solved instances per time
           for \textit{clasp~2} and other multi-threaded SAT solvers.}
  \label{fig:other}
\end{figure}%
% ------------------------------------------------------------

Our first series of experiments evaluates the performance of \textit{clasp}
in comparison to other multi-threaded SAT solvers.
To this end, we ran the aforementioned solvers on 160 benchmark instances from the Crafted category
at the 2011 SAT competition.%
\footnote{%
From the whole collection of 300 competition benchmarks,
the 160 selected instances could be solved with \textit{ppfolio} \cite{roussel11a},
the (wall-clock time) winner in the Crafted category at the 2011 SAT competition,
within 1000 seconds.
Without this preselection,
plenty (more) runs of the considered solvers would not finish in the time limit,
and running the experiments would have consumed
an order of magnitude more time.}
The plot in Figure~\ref{fig:other} displays numbers of solved instances (on the y-axis)
as a function of time (in log scale on the x-axis).
As (sequential) baseline,
we include \textit{clasp} running one thread in the configuration %
% \footnote{See \url{http://www.cril.univ-artois.fr/SAT11/results/solverlist.php?idev=46#solv1917}.}
submitted to the 2011 SAT competition.
This configuration is contrasted with four- and eight-threaded 
variants of the considered parallel SAT solvers,
using a prefabricated portfolio (\texttt{clasp {-}-create-template})
for competing threads of \textit{clasp}.
First of all, we observe in Figure~\ref{fig:other} that
all multi-threaded solvers complete more instances than sequential \textit{clasp}
when given sufficient time (more than 10 seconds).
This is unsurprising because the available CPU time roughly amounts
to the product of wall-clock time and number of threads,
given that our benchmark machine offers sufficient computing resources
for concurrent thread execution.
In fact, we further observe that each multi-threaded solver benefits from running more
(eight instead of four) threads.
However,
the increase in the number of solved instances is solver-specific and
rather small with \textit{manysat},
which mainly duplicates its fixed portfolio of four configurations
in the transition to eight threads (changing only the random seed used in the branching heuristics).
Unlike this, the other multi-threaded solvers complete between five (\textit{clasp})
and eight (\textit{cryptominisat}, \textit{miraxt}, and \textit{plingeling}) more instances 
in the time limit when doubling the number of threads.
These improvements are significant because harnessing additional computing
resources for parallel search is justified when it makes instances accessible
that are hard (or unpredictable) to solve sequentially.\footnote{%
The speedup (in terms of wall-clock time) of eight-threaded over single-threaded \textit{clasp}
is about 1.5, which may seem low,
but the eight-threaded variant completes 31 instances 
(with unknown sequential solving time) more.}
Comparing the performance of multi-threaded \textit{clasp} to other
SAT solvers shows that \textit{clasp} is very competitive, thus emphasizing the (low-level)
efficiency of its parallel infrastructure.
But please take into account that Crafted benchmarks are closer to ASP problems,
which \textit{clasp} is originally designed for, than those in
SAT competitions' Application category, to which the other four  SAT solvers
are tailored.
Finally, although solver portfolios (as used in \textit{ppfolio}) proved to be
powerful at the 2011 SAT competition,
we do  not include them in our experiments because their diverse members
are run     in separation, thus not utilizing multi-threading for parallelization.%
% ------------------------------------------------------------
\begin{figure}[t]
  \centering%\hspace*{-8mm}
  \includegraphics[width=0.5\textwidth]{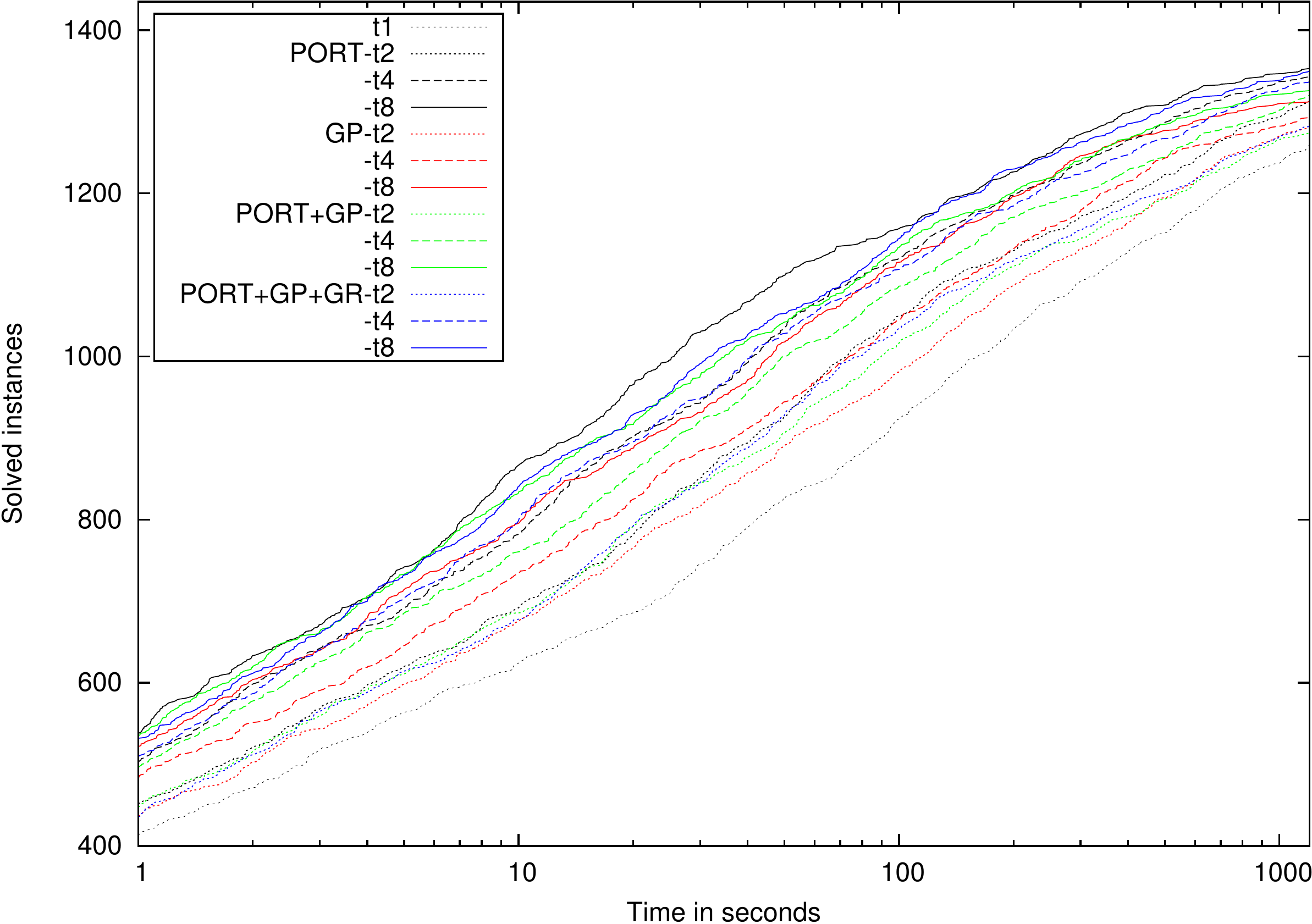}%
\skipsemi\skipsemii%
  \caption{Number of solved instances per time
           for different parallel search strategies of \textit{clasp~2}.}
  \label{fig:search}
\end{figure}%
% ------------------------------------------------------------

The second series of experiments assesses parallel search features of \textit{clasp}
on a broad collection of 1435 benchmark instances,
stemming from the 2009 ASP and SAT competitions as well as the 2006 and 2008 SAT races.
To begin with, the plot in Figure~\ref{fig:search} compares different parallel search strategies, viz.\
portfolio of competing threads (PORT), search space splitting via guiding paths (GP),
splitting-based search with a portfolio of different configurations (PORT+GP), and the
previous setting augmented with global restarts (PORT+GP+GR).
Note that the PORT mode matches the \textit{clasp} setup that has already been used above,
and that up to ten restarts (according to the geometric policy $500{*}1.5^i$)
are performed globally with the PORT+GP+GR mode.
As in our first experiments,
we observe that all multi-threaded \textit{clasp} modes dominate the
baseline of running a single thread.
Similarly, each mode benefits from more threads, where the transition from two to four
threads is particularly significant with portfolio approaches
(e.g.\ 32 more instances completed with PORT).
In fact, the latter dominate the GP mode relying on a uniform \textit{clasp}
(default) configuration, especially when the number of threads is greater than two.
This indicates the difficulty of making fair splits in view of irregular search spaces,
while running different configurations in parallel improves the chance of success
(cf.\ \cite{hyjuni11a}).
Although the robustness of splitting-based search is somewhat enhanced 
by running different configurations (PORT+GP) and additionally
applying global restarts to refine uninformed splits (PORT+GP+GR),
its combinations with guiding paths could not improve over the plain PORT mode.
However, it would be interesting to scale this experiment further up
(on a machine with more than eight cores) in order to investigate
whether a portfolio becomes saturated at some point, so that combinations
with search space splitting would be natural to % make use of 
exploit greater parallelism.%
% ------------------------------------------------------------
\begin{figure}[t]
  \centering%\hspace*{-8mm}
  \includegraphics[width=0.5\textwidth]{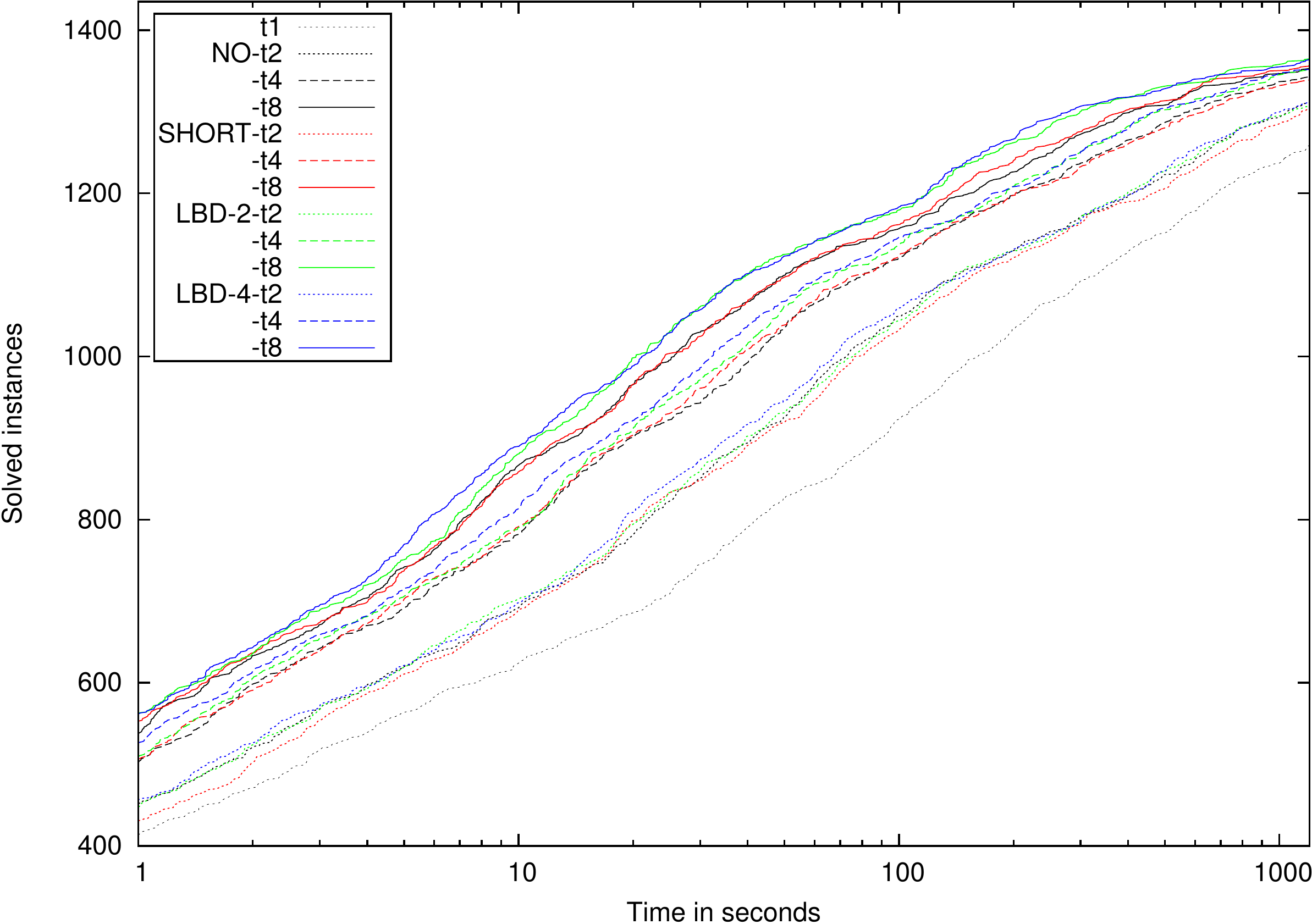}%
\skipsemi\skipsemii%
  \caption{Number of solved instances per time
           for different nogood exchange policies of \textit{clasp~2}.}
  \label{fig:dist}
\end{figure}%
% ------------------------------------------------------------

Finally, Figure~\ref{fig:dist} plots the performances of \textit{clasp} (PORT mode)
w.r.t.\ nogood exchange policies.
Given that the binary and ternary implication graph is always shared among all threads,
the difference between the NO and SHORT modes is that short nogoods are recorded 
``silently'' with NO and proactively communicated with SHORT
(cf.\ Section~\ref{subsec:nogood}).
The LBD-2 and -4 modes further extend SHORT by additionally distributing ``long'' nogoods
whose LBD does not exceed~2 or~4, respectively,
independent of the nogood size in terms of literals.
While the amount of solved instances is primarily influenced by the number of threads,
different nogood exchange policies are responsible for gradual differences between
\textit{clasp} variants running the same number of threads.
With four and eight threads, the LBD modes are more successful than NO and SHORT,
especially in the time interval from 10  to a few hundred seconds.
This shows that the exchange of information helps to reduce redundancies between
the search processes of individual threads;
it further supports the conjecture in \cite{audsim09a} that
``our measure [LBD] will also be very useful in the context of parallel SAT solvers.''
Interestingly, even when running eight threads,
the performances of LBD-2 and -4 modes are close to each other,
with a slight tendency towards LBD-4.
Our experiments do thus not exhibit bottlenecks 
due to the additional exchange of nogoods with LBD~3 and~4.
However, more exhaustive experiments are required (and part of our ongoing work)
to find a good trade-off between number of threads and LBD limit for exchange.
Ultimately, dynamic measures like those suggested in \cite{hajasa09b} are
indispensable for self-adapting nogood exchange to different problem characteristics, and
adding such measures to \textit{clasp} is a subject to future work.

%% file: relwork.tex
\section{Related Work}
\label{sec:relwork}

Parallel ASP solving was so far dominated by approaches distributing
tree search by extending the     solver \textit{smodels} in various ways
\cite{fimamotr01a,hirsimaki01a,pobabe03a,bapoelle05a,grjamescthti05a,grjamescthti06b}.
While
\textit{smodels} applies systematic backtracking-based search, following the scheme of 
DPLL
% the Davis-Putman-Logemann-Loveland (DPLL) procedure \cite{davput60,dalolo62a}
used in traditional SAT solving,
% Unlike this, 
\textit{clasp} as well as modern SAT solvers are based on 
CDCL,
% Conflict-Driven Clause Learning (CDCL) \cite{marsak99a,zamamoma01a},
relying on conflict-driven learning and backjumping.
However,
the clear edge of CDCL-based solvers over DPLL-based ones also brings about more
sophisticated search procedures that have to be accommodated in a distributed setting.
Apart from distributed constraint learning,
this particularly affects the coordination of model enumeration.

The approach taken with \textit{claspar} \cite{elgegukakaliscscsc09a,gekakascsc11a} 
can be regarded as a precursor to our present work.
\textit{claspar} is designed for a cluster-oriented setting without any shared memory.
% As such, it 
It thus aims at large-scale computing environments,
where physical distribution necessitates
% relying on 
data copying rather than sharing.
In fact, \textit{claspar} can be understood as a wrapper controlling the
distribution of independent \textit{clasp} instances via MPI \cite{grluth99a},
thereby taking advantage of \textit{clasp}'s interfaces for data exchange.
However,    compared   to \textit{claspar},
(quasi) instantaneous communication via shared memory
enables a much  closer collaboration (e.g.\ rapid nogood exchange) among
   threads in \textit{clasp}.

Although much work has also been carried out in the area of parallel logic programming,
among which or-parallelism \cite{gupoalcahe01a,kercod94a} is similar to search space splitting,
our work is      more closely related to % distributed
parallel SAT solving,
tracing back to \cite{zhbohs96a,blsiku03a}.
Among     modern approaches to multi-threaded SAT solving,
the ones of \textit{miraxt} \cite{sclebe09a} and \textit{manysat} \cite{hajasa09c}
are of particular interest due to their complementary treatment of recorded nogoods.
\textit{miraxt} is implemented via \emph{pthreads}
and uses a globally shared nogood database. % of learned clauses.
The advantage of this is that each thread sees all nogoods % learned clauses
and can integrate them
% their integration has thus a 
with low latency.
However, given that multiple threads read and write on the database,
it needs readers-writer locks.
Moreover,
many % clauses
nogoods are actually never used by more than one thread,
% leading to memory overhead.
but still produce some maintenance overhead in each thread.
% Similarly,
% threads have to visit all clauses even though most will not be integrated,
% resulting in cache overhead.
%
\textit{manysat} is implemented via \textit{openmp}
and uses a copying approach to nogood exchange,
proscribing any physical sharing.
That is, each among~$n$ solver threads has its own nogood database,
and nogood exchange is accomplished by copying via
$n{*}(n{-}1)$ pairwise distribution queues.
% nogoods by copying them via $n{*}(n{-}1)$ pairwise distribution queues.
While this approach performs well for a small number~$n$ of solver threads,
it % fails to scale with an increasing number of threads 
does not scale up due to the quadratic number
of queues and excessive copying.
Recent  parallel SAT solvers further include \textit{plingeling} \cite{biere11a} and
the multi-threaded variant of \textit{cryptominisat} \cite{sonoca09a}.
Finally, note that, while knowledge exchange and (shared) memory access matter likewise
in parallel SAT and ASP solving,
the scope of the latter also stretches out over % parallel 
enumeration and
optimization of answer sets.

%%% Local Variables: 
%%% mode: latex
%%% TeX-master: "paper"
%%% End: 

%% file: discussion.tex
\section{Discussion}\label{sec:discussion}

We have presented major design principles and key implementation techniques
underlying the \textit{clasp~2} series,
%To our knowledge, \textit{clasp~2} is 
thus providing the first CDCL-based ASP solver
supporting parallelization via multi-threading.
While its multi-threaded variant aims at leveraging the power of
today's multi-core shared memory machines in parallel search,
\textit{clasp~2} has also been designed with care 
not to sacrifice the 
(low-level) performance of its single-threaded variant,
sharing a common code base. 
In fact, the competitiveness of single- as well as multi-threaded \textit{clasp~2}
variants is,
for instance, witnessed by their performances at
the 2011 SAT competition.
Beyond powerful parallel search,
multi-threaded \textit{clasp~2} allows for conducting the various reasoning modes
of its single-threaded sibling,
including enumeration and (hierarchical) optimization, in parallel.
On the one hand, this makes the multi-threaded variant of \textit{clasp~2}
highly flexible, offering parallel solving capacities for various reasoning tasks.
On the other hand, the vast configuration space of a CDCL-based solver
becomes even more complex, as individual threads as well as their interaction
can be configured in manifold ways.
In view of this, 
adaptive solving strategies (e.g.\ regarding nogood exchange) and automatic % support for
parallel solver configuration are important issues to future work.

% \begin{itemize}
% \item first parallel CDCL-based ASP solver
% \item goes beyond parallel SAT in offering complex reasoning modes, like various
%   forms of enumeration and optimization
% \item Benjamin wins competition
% \item 
% In fact, it provides the first multi-threaded version of an ASP solver using
% conflict-driven learning.
% Unlike this, existing distributed ASP solvers rely on classical backtracking schemes
% that provide a much rigid traversal of the search space.
% Second, our approach endows each solver instance with great independence,
% principally allowing for a variety of different ASP solvers at the same time.
% \end{itemize}

%%% Local Variables: 
%%% mode: latex
%%% TeX-master: "paper"
%%% End: 

%% file: acknowledgments.tex
\paragraph{Acknowledgments}
We are grateful to Hannes Schr\"oder for support with experiments and to the
anonymous referees for their comments.
This work was partially funded by the German Science Foundation (DFG) under
grant SCHA 550/8-2.

%%% Local Variables: 
%%% mode: latex
%%% TeX-master: "paper"
%%% End: 

%% file: bbl.tex
%%% Local Variables: 
%%% mode: latex
%%% TeX-master: "paper"
%%% End: 